# CBIM: A Graph-based Approach to Enhance Interoperability Using Semantic Enrichment

Zijian Wang[1], Huaquan Ying[1], Rafael Sacks[1], André Borrmann[2]
[1]Technion Israel Institute of Technology, Israel
[2]Technical University of Munich, Germany
zijian.wang@campus.technion.ac.il

**Abstract.** Interoperability remains a challenge in the construction industry. In this study, we propose a semantic enrichment approach to construct BIM knowledge graphs from pure building object geometries and demonstrate its potential to support BIM interoperability. Our approach involves machine learning and rule-based methods for object classification, relationship determination (e.g., hosting and adjacent) and attribute computation. The enriched results are compiled into a BIM graph. A case study was conducted to illustrate the approach for facilitating interoperability between different versions of the BIM authoring software Autodesk Revit. First, pure object geometries of an architectural apartment model were exported from Revit 2023 and fed into the developed tools in sequence to generate a BIM graph. Then, essential information was extracted from the graph and used to reconstruct an architectural model in the version 2022 of Revit. Upon examination, the reconstructed model was consistent with the original one. The success of this experiment demonstrates the feasibility of generating a BIM graph from object geometries and utilizing it to support interoperability.

## 1. Introduction

The Architecture, Engineering, and Construction (AEC) industry is characterized by its fragmented structure, requiring collaboration among numerous stakeholders with specialized skills throughout a project's lifecycle (Borrmann *et al.*, 2018). Building Information Modelling (BIM), which presents models in a digital format, significantly enhances planning, design, construction, maintenance, and other processes across the entire building lifecycle (Sacks *et al.*, 2018). However, interoperability remains a challenge when delivering and exchanging BIM models. Software vendors have developed proprietary data schemas used within their own software ecosystems, resulting in data and files created by one vendor's applications incompatible with other vendors.

To improve the interoperability between various applications, the AEC industry has invested efforts in constructing open standards for sharing BIM information across platforms, such as the Industry Foundation Classes (IFC) schemas (buildingSMART, 2018). This also requires BIM software to compile models from proprietary schemas into the open schema, which potentially causes information loss and errors during the conversion process (Pazlar and Turk, 2008; Venugopal, Eastman and Teizer, 2015).

Semantic enrichment (SE) for BIM models has been proposed to infer implicit semantic information based on existing BIM data and supplement it back into the models (Belsky, Sacks and Brilakis, 2016; Bloch, 2022). A significant application domain of SE is to improve BIM data quality and enhance BIM interoperability. However, prior studies primarily concentrate on exploring techniques that can be employed on SE tasks, like adopting expert systems and deep learning algorithms to classify BIM objects. These studies validated the feasibility of various techniques for SE, but they presented the enriched results to users instead of inserting back to models (Bloch, 2022), and the prediction was limited to specific types of data instead of providing comprehensive required building information.

In this study, we explore adoption of SE tools to infer building semantics from object geometries and represent the results as a BIM graph. Subsequently, we propose a novel approach to enhance interoperability across software by leveraging the enriched semantics and software APIs. The launch point of SE is object geometries, which are the fundamental data of BIM models from LOD 200. Almost all BIM design software supports consistent export of solid geometries into generic and accessible file formats. Additionally, graphs have a flexible data structure and the inherent advantage of representing object relationships explicitly.

## 2. Background

Interoperability refers to the ability to exchange data across software and platforms, and the AEC industry continues to face challenges posed by interoperability issues (Sacks *et al.*, 2018). The fragmented nature of the AEC industry involves experts with a wide range of skills, leading that data to be represented in various formats to fulfill specific requirements. This diversity in data representation increases the complexity of interoperability. Furthermore, vendors create their own proprietary schemas for storing and exchanging data within their software ecosystems, resulting in the situation where files generated by one vendor's applications are potentially inaccessible on other platforms.

Researchers aim to use semantic enrichment to enhance interoperability. For example, Wu & Zhang (2019) proposed a nine-step rule-based approach to classify IFC entities into predefined categories, in which the classification scope is more detailed than the general object classes in the IFC schema, such as a U-shaped beam. Some researchers employ machine learning algorithms to re-identify IFC object types that may have been misclassified during the data exchange phase (Koo and Shin, 2018). In these cases, an SE process serves as an automatic data quality check.

Later, 3D geometric deep learning networks, such as Multi-view CNN and PointNet, were employed to predict subtypes of BIM door elements (Koo, Jung and Yu, 2021). Furthermore, solid BIM entities were sampled into points and converted into an object graph to enable the use of graph neural networks (GNNs) (Collins *et al.*, 2021). Researchers also constructed apartment BIM graphs, with nodes representing spaces and edges denoting relationships, and implemented a customized GNN algorithm to classify space types (Wang *et al.*, 2021; Wang *et al.*, 2022). In addition, Costa & Sicilia (2020) explored the potential of representing IFC models as RDF graphs to enable semantic web query and facilitate information exchange across platforms.

These studies have expanded the scope of techniques available for SE. However, they narrowly focus on developing techniques for individual SE tasks. As a result, the enrichment is limited to specific types of information, such as building object types. Instead, the missing information causing interoperability issues goes beyond object types and encompasses relationships between objects, object attributes, and even more. A comprehensive approach that can generically enrich various types of information is a necessary pre-requisite to address interoperability issues.

In addition to the lack of semantics, Bloch (2022) also points out that the predicted data are often presented to users rather than being integrated back into models, leading to difficulties in reusing the data in downstream applications. To address this problem, researchers have started to adopt graphs to represent building information (Venugopal, Eastman and Teizer, 2015; Pauwels and Terkaj, 2016). A graph consists of nodes and edges with the advantage of storing information flexibly and representing relationships explicitly (Zhou *et al.*, 2018). By leveraging these explicit relationships, researchers have developed various applications on building graphs, such as searching shortest fire egress path (Ismail, Strug and Ślusarczyk, 2018), querying information (Khalili and Chua, 2015) and semantic enrichment (Wang, Sacks and Yeung, 2022). Furthermore, the Linked Building Data (LBD) community has made efforts to standardize the terminologies and formats of BIM graphs by constructing ontologies and using the techniques from the semantic web domain (Pauwels, Costin and Rasmussen, 2022).

The primary objective of this study is to explore a new approach that uses multiple SE tools to construct a BIM graph for addressing the interoperability issues. To achieve this, we: 1) propose a pipeline to construct a BIM knowledge graph by implementing SE tools on the fundamental object geometries of BIM models; 2) design an approach to ease the interoperability problem by reconstructing BIM models across software by leveraging BIM graphs.

## 3. Pipeline of Constructing BIM Graphs from Object Geometries

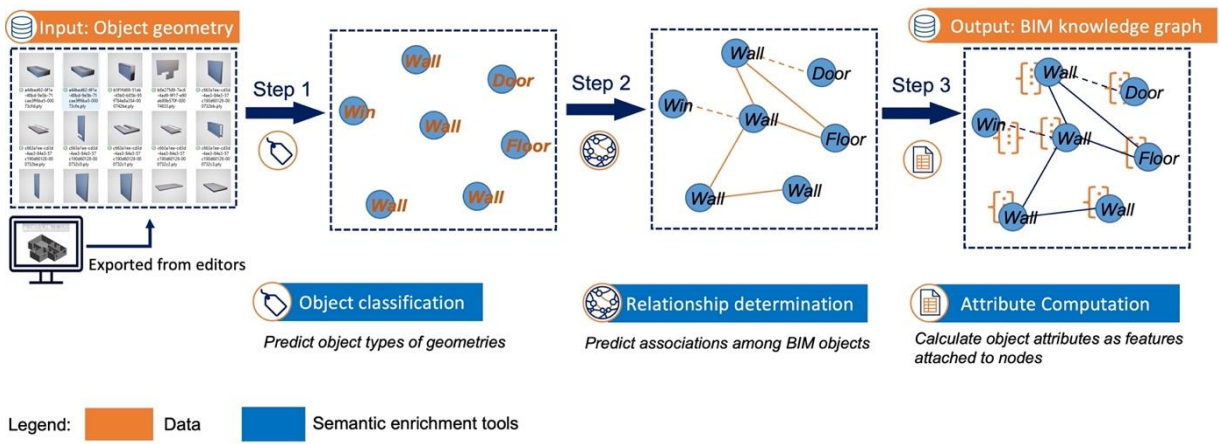

Figure 1: Pipeline of generating BIM knowledge graphs from object geometries.

Figure 1 presents the pipeline using SE tools to construct the BIM knowledge graph from object geometries. It employs three artificial intelligence tools to infer BIM semantics in sequence and to represent the inferred semantics as an interoperable graph. The input, pure object geometries, can be exported from BIM editors and stored in a generic file format, such as PLY, which represents geometries as triangle or polygon meshes in a global coordinate system. Additionally, the names and usages of nodes and edges in the knowledge graph follow ontologies from the LBD community to ensure its accessibility. Tools from the semantic web domain, like RDF, are adopted to compile and retrieve graphs.

To design experiments for validating this pipeline, the case of apartment design at the initial design stage was selected. Architects frequently exchange model files with participants and sometimes reconstruct their models from scratch. From the perspective of implementation, this scenario is not overly complex for evaluating the feasibility. Specifically, the first step in the enrichment process is to classify the object types from geometries (detailed in Section 3.1), where the predicted results are inserted back into the graph (as denoted by the first graph database of Figure 1). Subsequently, hosting or hosted and adjacent relationships among BIM objects are instantiated by leveraging geometry features, as detailed in Section 3.2. Lastly, attributes are computed based on geometries and object types, and stored as node features in the graph (see Section 3.3).

### 3.1. Object classification

Classifying BIM objects is a well-studied problem, with numerous approaches investigated by researchers. Wu et al. (2022) discovered that machine learning algorithms utilizing geometric, locational, and metadata features can achieve accurate and robust performance, with around 99% precision for five classes. However, machine learning often surpasses deep learning in computational efficiency and provides greater scalability compared to rule-based algorithms, which require manual development of rule sets. Considering these factors, we adopted machine learning for the task of apartment object classification.

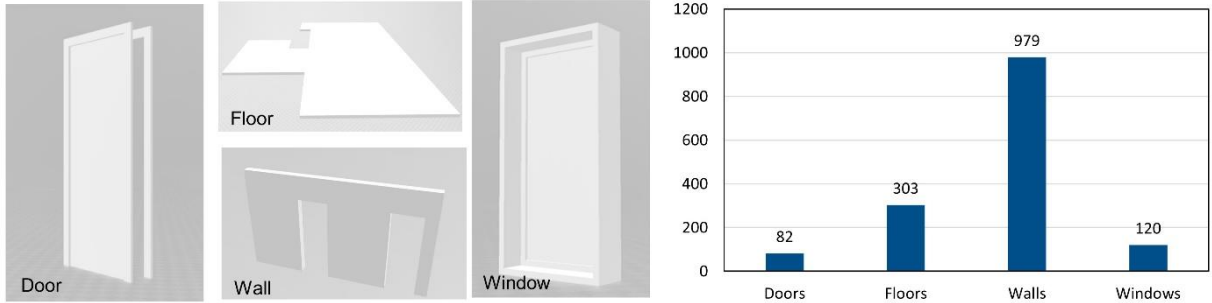

Figure 2: Dataset samples and statistics.

In the initial design phase of an apartment, BIM models typically include four object types: walls, floors, windows, and doors. We collected 32 architectural apartment units from real projects in Israel, which were originally designed in Revit 2019. Each architectural building object was then extracted from the Revit database and saved as a single PLY mesh geometry file using the Dynamo exporter (Wang, et al. 2023). The dataset comprises 1,484 object instances after removing duplicates, as shown in Figure 2.

Table 1: Results of nine machine learning algorithms on the constructed datasets.

| No. | Machine learning algorithm | Valid accuracy | Test accuracy |
| --- | --- | --- | --- |
| 1 | Logistic regression | 65.79% | 62.45% |
| 2 | Support vector machine (SVM) | 68.71% | 69.80% |
| 3 | K-nearest neighbors (KNN) | 89.54% | 77.96% |
| 4 | Gaussian naive bayes | 74.95% | 74.90% |
| 5 | Perceptron | 20.52% | 20.20% |
| 6 | Linear support vector classification | 45.47% | 46.33% |
| 7 | Stochastic gradient descent (SGD) | 20.52% | 20.20% |
| 8 | **Decision tree** | **100.00%** | **100.00%** |
| 9 | **Random forest** | **100.00%** | **100.00%** |

Nineteen features related to object dimensions, mesh characteristics, and locations were utilized in the feature engineering, such as area, volume, number of faces, maximum value on the x-axis, and so on. The dataset was randomly divided into training, validation, and test sets at a ratio of 7:1:2. Nine classic machine learning algorithms were implemented using Python and the scikit-learn library. The results are displayed in Table 1. Both the decision tree and random forest algorithms demonstrate a strong ability to classify objects, achieving 100% accuracy for both the validation and test sets. This indicates that these two algorithms can effectively predict the types of geometries. We selected random forest as the algorithm for apartment object classification in this study.

### 3.2. Relationship determination

This study aims to generate two types of object relationships. The first is adjacency, which describes the topological fact that two objects are physically touching each other. We use cbim:adjacentTo from the CBIM ontology to represent the generated adjacency relationships (Sacks *et al.*, 2022). The second type is hosting, a semantic relationship illustrating that one object functionally hosts another. We adopt cbim:hosting/cbim:hosted from the CBIM ontology to display the hosting relationships with directions.

We designed a rule-based method to generate hosting, hosted and adjacent relationships by leveraging object geometry characteristics. First, we calculate the bounding box distance in three axes for two objects and proceed only if the sum of the three distances is zero. This step filters out pairs of objects that cannot physically touch and speeds up the process. Then, we compute the exact geometry distance between the two objects and continue only if the distance is close to zero, which avoids the situations where two objects are either too far apart or overlapping. After that, we use the topological relationships of the two objects' bounding boxes in three dimensions to determine the final relationship. For example, if the bounding box of one object contains the bounding box of another object, we assign the cbim:hosting relationship. Similarly, if one object's bounding box is consistently contained by the bounding box of another object, we assign a cbim:hosted instance. Otherwise, cbim:adjacentTo is assigned. Experiments show that the method can achieve an accuracy and F1 score of 0.99 on test data.

### 3.3. Attribute computation

In this study, we primarily compute dimensional attributes of objects using object geometries and object types. Geometry serves as the basic source for measurement, and object types assist in determining attribute names. Taking the slab as an example, the measure in the Z-dimension can be interpreted as thickness. However, when evaluating a wall, the measure value in the Z-dimension would be regarded as height. The algorithms developed in this study can compute various attributes, such as area, volume, height, width, perimeter, thickness, slope, length, and depth.

### 4. Applications on Enhancing Interoperability Using BIM Graphs

Designers frequently encounter situations where they cannot access models saved in one software's default format when using another vendor's software. Sometimes, even within the same vendor environment, earlier version software is unable to read a model created by later version software. For example, legacy versions of Revit cannot access models created by the latest version of Revit. To address the interoperability issue, users often need to export models from the proprietary schema to an open data schema like IFC.

In this study, we propose a novel approach to alleviate the interoperability problem by leveraging the generated BIM graph and software APIs. The process of downgrading a Revit model to an earlier version is illustrated in Figure 3. The whole process mainly contains three parts: 1) Exporting object geometries from a later version Revit 2023; 2) Generating the BIM graph by the developed SE tools, 3) Reconstructing BIM models in Revit 2022 using the enriched semantics and software APIs.

In the first step of Figure 3, object geometries were exported by a program from the Revit 2023 database and saved as independent PLY files. The Geometry exporter is implemented by Dynamo (Wang et al., 2023), which leverages the APIs released by Autodesk.

Next, the three SE tools were executed in sequence to predict the object types, instantiate hosting, hosted and adjacent relationships, and calculate object attributes. All the predicted results were organized into a knowledge graph, as shown in Figure 4. Nodes in the graph represent BIM objects, each named by their object type and unique ID. Edges indicate object-object relationships. The first part of Figure 4 illustrates the cbim:adjacentTo relationships between two walls, while the second part showcases the cbim:hosted or cbim:hosting relationships between walls and windows or doors. Attributes are attached to nodes but hidden in the graph for readability.

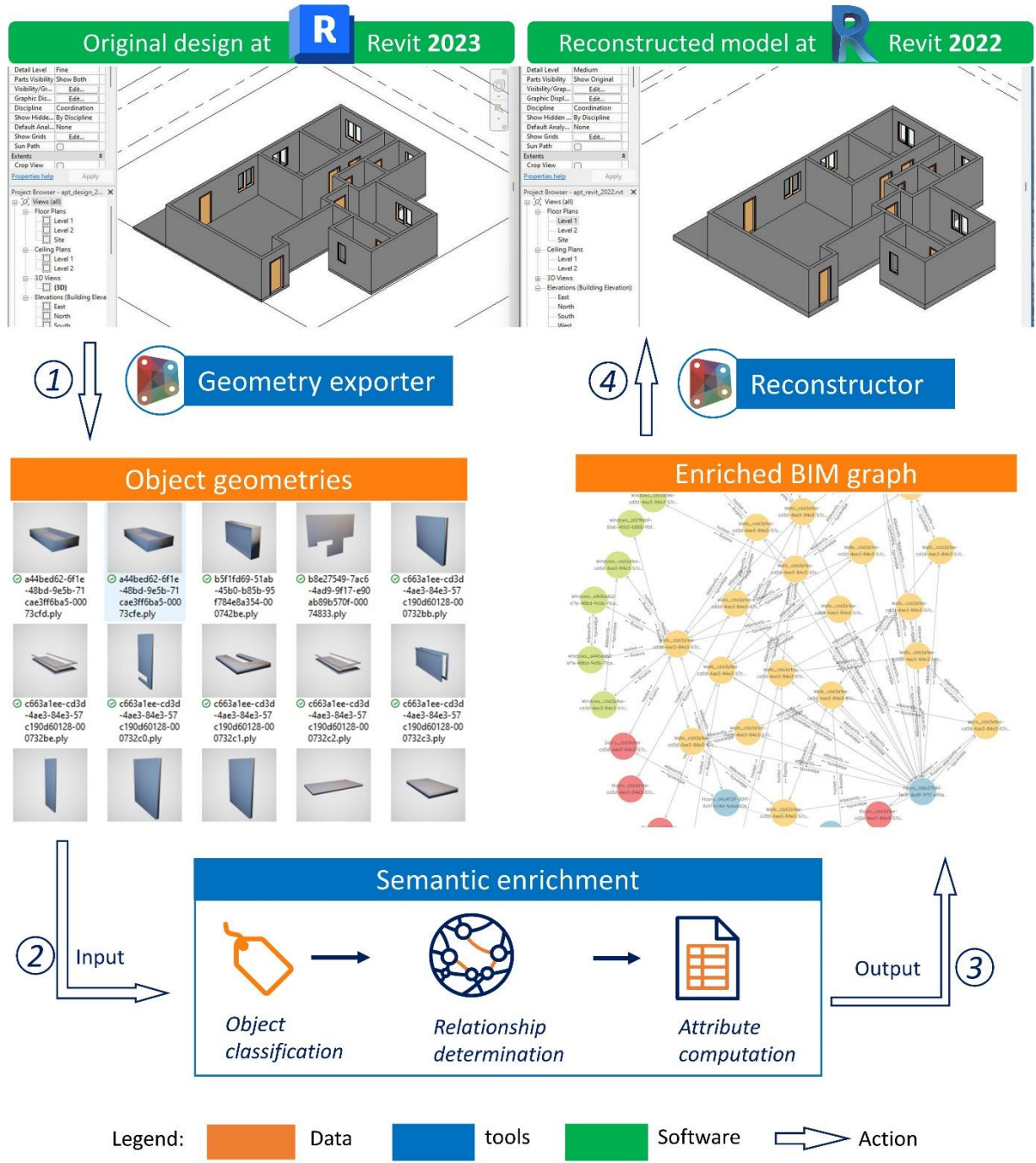

Figure 3: Process of using semantic enrichment to improve interoperability: a Revit downgrading case.

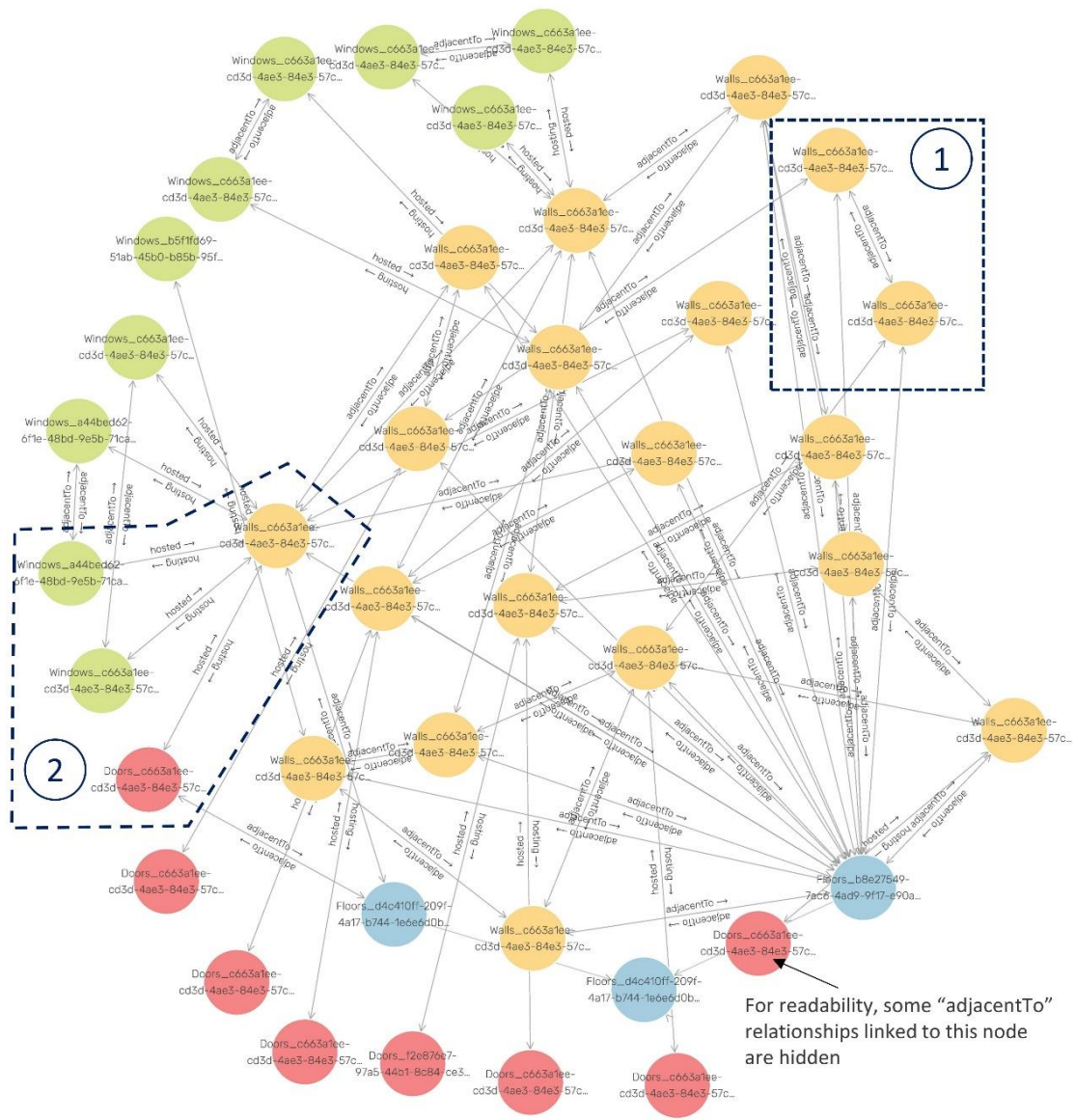

Figure 4: Generated BIM knowledge graph of an architectural apartment model
(① illustrates the cbim:adjacentTo relationships, ② depicts the cbim:hosting/hosted relationships).

Finally, a Reconstructor was employed to retrieve the necessary information from the BIM graph and construct objects in a lower version of the software, Revit 2022. The Reconstructor is implemented in Dynamo, and part of the code is presented in Figure 5. Generally, it reads the knowledge graph by object types and extracts the information required for object construction. Taking the window construction as an example, it needs the hosted object, the window type, and the window's central point. Therefore, the Reconstructor begins by reading one window instance and searching for its hosted wall using the explicit cbim:hosted relationship. Additionally, it retrieves the window's locations and computes the central point. The reconstruction process utilizes all enriched semantics, including object types, hosted relationships, and computed attributes.

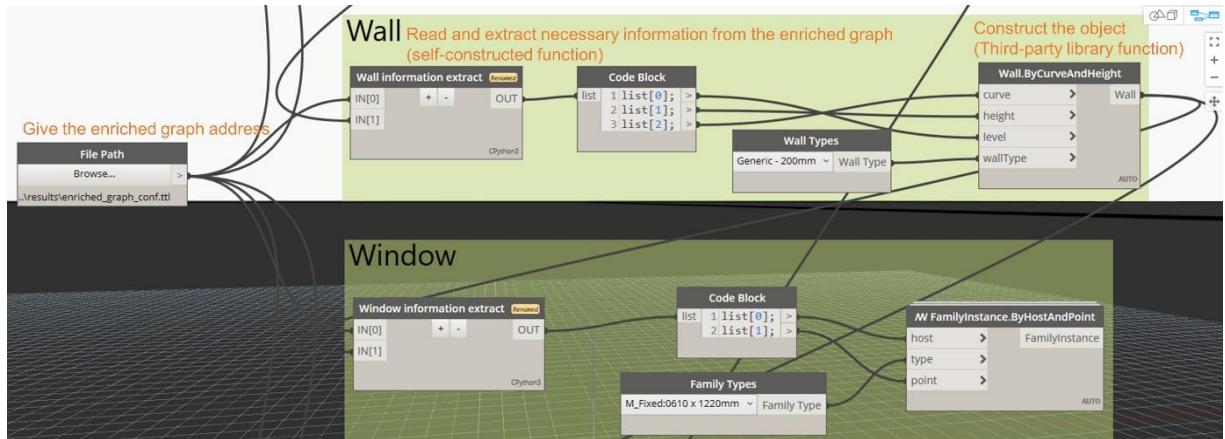
Figure 5: Diagram of part of the Dynamo code from the Reconstructor.

We performed a manual comparison of each object pair across the two different Revit models. We found that the location, dimensions, and object types matched. It is worth noting that the object IDs have been changed, as this process focuses on reconstructing models rather than duplicating information. Furthermore, one requirement for successful reconstruction in Revit 2022 is that Dynamo should have permission to access the internal schema and modify the data in Revit's internal database.

## 5. Discussion

The first strength of this study is the proposed novel approach of generating BIM graphs from object geometries and using the knowledge graph to downgrade design models from a later version to an earlier one, addressing interoperability issues. Furthermore, we developed a Dynamo Reconstructor program to retrieve information on graphs and generate objects within the Revit environment.

Despite these strengths, this study faces several limitations, which can guide future research directions. Firstly, there is still room for enriching BIM graphs. For example, building spatial structural objects, from the site to the building to the levels, represent essential hierarchy information (Ying *et al.*, 2019). Additionally, space objects are necessary for running various building performance simulations (Ying and Lee, 2021) and quantity take-offs (Bloch and Sacks, 2018).

Secondly, this study achieves three types of SE tasks: object classification, relationship determination and attribute computation. The need remains to explore object generation, another type of SE task (Bloch and Sacks, 2020). One direction could involve leveraging physical building objects to generate virtual objects automatically, such as using walls and floors to create space objects. This could reduce manual modeling efforts in the design phase and provide useful semantics for downstream applications.

Thirdly, this study showcases the use of enriched graphs for interoperability, but there is still a need to develop more intelligent applications on graphs. Researchers have investigated the feasibility of controlling design versions on graphs (Esser, Vilgertshofer and Borrmann, 2022) and demonstrated interdisciplinary consistency maintenance (Wang, Ouyang and Sacks, 2022).

## 6. Conclusion

This study explores the adoption of semantic enrichment to generate BIM graphs for enhancing the interoperability of BIM. The main conclusions and findings are as follows:

- We proposed a pipeline for predicting various types of semantics from object geometries and organizing them as a knowledge graph.
- We illustrated a novel approach of using the BIM graph to enhance interoperability by downgrading a later version Revit model to an earlier version model.

The success of the experiment demonstrates the feasibility of constructing building knowledge graphs from object geometries to improve the interoperability of BIM. This research can pave the way for researchers to develop more SE tools using object geometries and explore more intelligent applications on BIM graphs.

## Acknowledgements

The work is part of the Cloud-based Building Information Modelling (CBIM) project, a European Training Network. The CBIM project receives funding from the European Union's Horizon 2020 research and innovation programme under the Marie Skłodowska-Curie grant with agreement No 860555.